\documentclass{iau} 
\usepackage{graphicx}

\title[Resolved stellar mass buildup and quenching in massive disk galaxies] 
{Spatially resolved stellar mass buildup and quenching in massive disk galaxies over the last 10 Gyr revealed with spatially resolved SED fitting}

\author[Abdurro'uf \& Masayuki Akiyama]   
{Abdurro'uf$^{1,2}$
 \and Masayuki Akiyama$^1$}

\affiliation{$^1$Astronomical Institute, Tohoku University, Aramaki, Aoba, Sendai 980-8578, Japan \\ email: {\tt abdurrouf@astr.tohoku.ac.jp} \\[\affilskip]
$^2$Institute of Astronomy and Astrophysics, Academia Sinica, Taipei 10617, Taiwan}

\pubyear{2018}
\volume{341}  
\jname{PanModel2018: Challenges in panchromatic galaxy modelling with next generation facilities}
\editors{M. Boquien, E. Lusso, C. Gruppioni, \& P. Tissera, eds.}
\begin{document}

\maketitle

\begin{abstract}
Despite decreasing cosmic star formation rate density over the last 10 Gyr, the stellar mass ($M_{*}$) buildups in galaxies were still progressing during this epoch. About 50\% of the current $M_{*}$ density in the universe was built over the last $\sim 8.7$ Gyr. In this research, we investigated the stellar mass buildup and quenching of spatially resolved regions within massive disk galaxies over the last 10 Gyr. We apply the spectral energy distribution (SED) fitting method to SEDs of sub-galactic regions in galaxies to derive the spatially resolved distributions of SFR and $M_{*}$ in the galaxies. This namely \textit{pixel-to-pixel SED fitting} method is applied to massive disk galaxies at $0.01<z<0.02$ and $0.8<z<1.8$. We found that massive disk galaxies tend to build their $M_{*}$ and quench their star formation progressively from the central region to the outskirts, i.e. \textit{inside-out} stellar mass buildup and quenching.  
\end{abstract}

\firstsection 
\section{Introduction}
Galaxy redshift surveys over the last decades have revealed dozens of global galaxy scaling relations. One important scaling relation is a star formation main sequence (SFMS) which is a tight nearly linear relation between global star formation rates (SFRs) and stellar masses ($M_{*}$) of star-forming galaxies. This relation holds across a wide redshift range and evolves in terms of the slope and the normalization of its linear form.

The decreasing normalization of the SFMS with cosmic time (a factor of $\sim 2$ dex from $z\sim 6$ to $0$ based on \cite[Speagle et al. (2014)]{Speagle2014}) suggests a decreasing global specific SFR (sSFR$\equiv$SFR$/M_{*}$) which implies a decreasing rate of stellar mass buildup and star formation quenching in galaxies.  

Besides the SFMS, the galaxy distribution is also bimodal on an optical color (e.g. $u-g$) versus $M_{*}$ plane (i.e. color-mass diagram). In contrast to the early-type galaxies which predominantly reside in the red sequence on the color-mass diagram, late-type galaxies make a continuous population ranging from the blue cloud all the way to the red sequence, suggesting a slow quenching process in the late-type galaxies (\cite[Schawinski et al. (2014)]{Schawinski2014}). A slow quenching in late-type galaxies could be caused by a cut-off of cosmic gas supply (i.e. strangulation) into the galaxy or inefficient gas cooling due to some feedback mechanisms (\cite[Schawinski et al. (2014)]{Schawinski2014}). Once the cosmic gas supply of late-type galaxy is cut-off, star formation still proceeds and accumulates stellar mass using the remaining gas until the gas in the galaxy is used up. Once the gas is used up, star formation will stop and the galaxy eventually become a red late-type galaxy.         

In this contributed paper, we discuss the spatially resolved distributions of SFR and $M_{*}$ in massive disk galaxies and study their structural evolution over the last 10 Gyrs. We discuss the gradual processes of star formation quenching and stellar mass buildup in those galaxies.

\section{Methodology}

In this research, we use a spatially resolved SED fitting method, i.e. SED fitting technique applied to spatially resolved SED of a galaxy. The reason for using this method over an analysis using the integral field spectroscopy (IFS) data is that this method is applicable to sample galaxies in the local universe and at high redshift with nearly the same manner. The usage of a single method can reduce some systematic biases in comparison analysis between results obtained with the two samples. We chose massive ($\log(M_{*})>10.5$) face-on disk galaxies at $0.01<z<0.02$ and $0.8<z<1.8$ as the sample of galaxies in this research. The local and high redshift samples contain 93 and 152 galaxies, respectively. The local and high redshift samples are drawn from the MPA-JHU and 3D-HST (for the GOODS-South field), respectively.  

The \textit{pixel-to-pixel SED fitting} method used in this research can be divided into three main tasks: (1) construction of maps of the multiband fluxes of a galaxy from which spatially resolved SEDs of the galaxy are obtained, (2) construction of a set of model photometric SEDs with a large number of random parameters. The model SEDs are redshifted according to the redshift of the sample galaxy, (3) fitting of a spatially resolved SED with a set of model SEDs using the Bayesian statistics approach. Detailed description of this method is presented in \cite[Abdurro'uf \& Akiyama (2017)]{Abdurrouf2017}. In order to construct spatially resolved SED with rest-frame far-ultraviolet (FUV) to near infrared (NIR) coverage, we use 7 bands from GALEX and SDSS for the local sample and 8 bands from CANDELS and 3D-HST for the high redshift sample. In order to get spatially resolved SEDs with high S/N ratio, we bin neighboring pixels that have similar SED shapes. The binning method is described in \cite[Abdurro'uf \& Akiyama (2017)]{Abdurrouf2017}. We require all the spatially resolved SEDs to have a S/N larger than 10 in all filters.

The model SEDs are generated using \cite[Bruzual \& Charlot (2003)]{Bruzual2003} stellar population synthesis (SPS) model assuming a \cite[Chabrier (2003)]{Chabrier2003} initial mass function (IMF), various stellar metallicities ($Z$) ranging from $0.004$ to $0.05$, exponentially declining star formation histories (SFHs), and \cite[Calzetti et al. (2000)]{Calzetti2000} dust extinction law. The Bayesian SED fitting that we adopted uses the Student's t distribution form for the likelihood function instead of Gaussian function. This likelihood function gives reasonable weight to the models with large $\chi^{2}$, thereby allowing us to explore the models over whole parameter ranges. We use a flat prior over an assumed range for all model parameters.        

\section{Results and discussion}

We investigated a linear scaling relation at a kpc scale between SFR surface density ($\Sigma_{\rm SFR}$) and $M_{*}$ surface density ($\Sigma_{*}$), the so-called spatially resolved SFMS. This relation holds in both local and high redshift samples (Abdurro'uf \& Akiyama 2017, 2018). Galaxies with lower global sSFR tend to have spatially resolved SFMS with lower normalization. Moreover, we found a tendency toward flattened spatially resolved SFMS relation in galaxies with lower global sSFR. Spatially resolved SFMS of massive star-forming (those residing on the global SFMS) disk galaxies at $0.8<z<1.8$ have a linear form with a slope of nearly unity over the entire $\Sigma_{*}$ range without any 'flattening' tendency in the high $\Sigma_{*}$ region (\cite[Abdurro'uf \& Akiyama (2018)]{Abdurrouf2018}). This trend suggests a similar rate of stellar mass buildup over the entire region in those galaxies. The 'flattening' tendency at high $\Sigma_{*}$ is observed in spatially resolved SFMS of galaxies with lower global sSFR than those of massive star-forming galaxies at $0.8<z<1.8$. The 'flattening' trend suggests that a quenching mechanism is taking place in sub-galactic regions at the high-mass end (predominantly located in the central regions of the galaxies). The spatially resolved SFMS relation shows that the amount of accumulated stellar mass and recent SFR trace each other even in a kpc scale within a galaxy. The origin of this relation has not been understood. Information on the spatially resolved gas surface density will be helpful to investigate the connection of this relation with the Kennicutt-Schmidt relation.      

In order to investigate the internal structure of the galaxies in the sample, we derived the radial profiles of the surface densities of SFR ($\Sigma_{\rm SFR}(r)$), $M_{*}$ ($\Sigma_{*}(r)$), and sSFR$(r)$. We found that the average sSFR$(r)$ of the massive star-forming (that reside on the global SFMS) disk galaxies at $0.8<z<1.8$ is roughly flat over the entire radial distances, suggesting a similar rate of stellar mass buildup over the entire region in these galaxies. The average sSFR$(r)$ of the high redshift sample that reside below the global SFMS is lower over the entire radial distances compared to the average sSFR$(r)$ of the star-forming galaxies with the central regions showing more significant suppression (\cite[Abdurro'uf \& Akiyama (2018)]{Abdurrouf2018}). The sSFR$(r)$ of all the local sample show suppression in the central region while roughly flat in the outskirts (\cite[Abdurro'uf \& Akiyama (2017)]{Abdurrouf2017}). Overall, decreasing global sSFR is associated with suppression of the sub-galactic sSFR over the entire galaxy region with a sharper suppression in the central region compared to that in the disk. Similar trends were also observed in local galaxies by \cite[Belfiore et al. (2018)]{Belfiore2018} and \cite[Abramson et al. (2014)]{Abramson2014}. This trend suggests that a quenching process in a galaxy is propagating from the central region to outskirts, i.e. \textit{inside-out quenching}. 

In order to study the evolution of the internal structure of massive disk galaxies in a more quantitative manner, we tried to construct an evolutionary empirical model using the surface density radial profiles of the local and high redshift samples. First, we connect the high redshift and local samples by looking for a possible pair of progenitors and descendants. We did that using an evolutionary track on the sSFR versus $M_{*}$ plane assuming an exponentially declining SFH with $\tau=[4.0:6.0]$ and an initial condition at $z=2$ with $\log(\rm sSFR (z=2))=[-8.6:-8.4]$ and $\log(M_{*}(z=2))=[9.7:9.9]$ drawn from the scatter of the global SFMS at $z=2$. Then progenitors and descendants are defined as galaxies in the high redshift and local samples that are passed by the model evolutionary track at the redshift of those galaxies, respectively. The number of progenitors and descendants are 20 and 14, respectively. The average $\Sigma_{\rm SFR}(r)$, $\Sigma_{*}(r)$, and sSFR$(r)$ radial profiles of the progenitors and descendants are then calculated. Using the average $\Sigma_{\rm SFR}(r)$ of the progenitors and descendants as a boundary condition and assuming an exponentially declining SFH at each radial distance, we estimated the radial profile of the SFH, $\Sigma_{\rm SFR}(r,t)$. The $\Sigma_{\rm SFR}(r,t)$ and average $\Sigma_{*}(r)$ of the progenitors form a set of formula for the evolutionary empirical model of the surface density radial profiles. The predicted $\Sigma_{*}(r)$ and sSFR$(r)$ at $z=0$ by the empirical model are consistent with the average $\Sigma_{*}(r)$ and sSFR$(r)$ of the descendants. Given that the empirical model is derived only using the average $\Sigma_{\rm SFR}(r)$ of progenitors and descendants, the consistency in $\Sigma_{*}(r)$ and sSFR$(r)$ implies that the empirical modeling does make sense.

The empirical model for the evolution of $\Sigma_{*}(r)$ shows stellar mass buildup in the disk, while stellar mass accumulation in the central region is small. The $\Sigma_{*}(r,t)$ shows stellar mass buildup progressing from the central region toward the outskirts, i.e. \textit{inside-out stellar mass buildup}. The $\Sigma_{\rm SFR}(r,t)$ and sSFR$(r,t)$ show decreasing SFR and sSFR with cosmic time over the entire radial distances with larger suppression in the central region compared to that in the outskirts. This trend suggests a quenching process that propagates from the central region to the outskirts, i.e. \textit{inside-out quenching}. However, the sSFR in the outskirts decreases mildly with cosmic time meaning that a quenching process also takes place in the outskirts. Using the empirical model and assuming $\log(\rm sSFR)=-10$ as a critical condition for quenching, we calculated the quenching timescale as a function of radius ($t_{\rm quench}(r)$). We found that the central region ($r\sim 1$ kpc) quenched by $\sim 200$ Myr from $z=1.1$, while the outskirts quenched by $\sim 5.2$ Gyr from $z=1.1$.

As suggested by $t_{\rm quench}(r)$, the quenching process progresses slowly in the disk. Although the disk slowly decreases its SFR, the stellar mass buildup is progressing in the disk. We see a simultaneous process of the \textit{inside-out growth} and \textit{inside-out quenching}. Both processes are connected in such a way that star formation was intense in the central region in the past (perhaps at the cosmic noon) but then quenched quickly. As star formation in the disk quenches slowly, the emerging picture is the stellar mass buildup that progresses \textit{inside-out}. 

These results agree with the slow quenching scenario of the late-type galaxies as suggested by \cite[Schawinski et al. (2014)]{Schawinski2014} and furthermore reveal the internal quenching process going on in the galaxies. It is possible that the disk galaxies observed in this research may experience a cut-off of cosmic gas supply (i.e. strangulation) since it can make SFR decreases slowly (perhaps exponentially) with time until the remaining gas in the galaxy is used up. The strangulation can happen once the halo mass reaches a critical mass ($\sim 10^{12}M_{\odot}$) above which a gas accretion to the halo enters a state of hot mode accretion (\cite[Birnboim \& Dekel (2003)]{Birnboim2003}). The strangulation is also suggested as a primary mechanism for quenching in galaxies by \cite[Peng, Maiolino, \& Cochrane (2015)]{Peng2015}. 

\section{Conclusion}

We investigated the spatially resolved distribution of $\Sigma_{*}$ and $\Sigma_{\rm SFR}$ in massive face-on disk galaxies at $0.01<z<0.02$ and $0.8<z<1.8$. We derived the spatially resolved $\Sigma_{*}$ and $\Sigma_{\rm SFR}$ in each galaxy using a method named \textit{pixel-to-pixel SED fitting}. In this method, spatially resolved SEDs of a galaxy are fitted to a set of model SEDs using Bayesian statistics approach. We found a more fundamental SFMS relation at a kpc scale which is a nearly linear relation between $\Sigma_{\rm SFR}$ and $\Sigma_{*}$. We investigated the $\Sigma_{\rm SFR}(r)$, $\Sigma_{*}(r)$, and sSFR$(r)$ radial profiles of the local and high redshift samples to study the structural evolution of massive disk galaxies and constructed an evolutionary empirical model. Overall, we observed simultaneous stellar mass buildup and quenching processes that progress \textit{inside-out}.

\begin{discussion}
\discuss{Tomo Goto}{MS flattening is happening to all sub-samples. How does it get steeper after integration?}
\discuss{Abdurro'uf}{The 'flattening' at high $\Sigma_{*}$ end in the spatially resolved SFMS relation is caused by the suppression of sSFR in the central regions of the galaxies. The larger normalization difference in the global SFMS relation given only slight normalization difference in the spatially resolved SFMS is caused by the inclusion of the bulge component when calculating integrated SFR. The bulge component gives large contribution to the integrated $M_{*}$ but only gives small contribution to the integrated SFR.}
\end{discussion}

\end{document}